\documentclass[twocolumn]{jpsj2tc} 
\title{Bayesian Nash Equilibria and Bell Inequalities}

\author{Taksu \textsc{Cheon}\thanks{Email address: taksu.cheon@kochi-tech.ac.jp} 
and 
Azhar \textsc{Iqbal}\thanks{Email address: azhar.iqbal@kochi-tech.ac.jp}}

\inst{Laboratory of Physics,
Kochi University of Technology,
Tosa Yamada, Kochi 782-8502, Japan
}

\abst{
Games with incomplete information are formulated in a multi-sector
probability matrix formalism that can cope with quantum 
as well as classical strategies.
An analysis of classical and quantum strategy in 
a multi-sector extension of the game of Battle of Sexes
clarifies the two distinct roles of nonlocal strategies,
and establish the direct link between the true quantum gain
of game's payoff and the  breaking of Bell inequalities. 
}

\kword{quantum Harsanyi game, nonlocality, entanglement, battle of sexes}

\begin{document}
\maketitle

\section{Introduction}

Although Bell's inequalities \cite{BE64,BE66} are usually
discussed in the context of quantum Bell experiments with spins and
observers, they can be established in a far wider variety of settings. Here
we bring one such example in a rather unexpected field of the theory of
games of incomplete information \cite{HA67}. Game theory now occupies a
central place in areas of applied mathematics, economics, sociology, and in
mathematical biology. It is well known that Bell inequality can be broken
only when the assumption of local realism is abandoned. This result, when
considered in the context of game theory of incomplete information, links
together the breaking of Bell inequality and the existence of nonlocal
correlation between players. To explore this further, a consideration of
quantum strategies \cite{ME99,EW99,OS04,CT06,FH07} in games of
incomplete information becomes both relevant and interesting.

With rapid advancement of quantum information technologies,
playing games with quantum resources is within the technical reach of
advanced laboratories \cite{DLX02,PSWZ07}. It is quite conceivable that
playing games with quantum strategies, using properly coordinated quantum
devices, becomes commonplace in the near future. 
It is therefore timely
that we analyze the physical contents of
quantum strategies, and examine the relevance of Bell inequality breaking.
%
It is now generally agreed 
that quantum strategy can shift the classical outcome of the game in favor of
all players, but how much of it is due to truly quantum effect, never
achievable classically, is still under debate \cite{IQ04}.
Games with incomplete information synergetic to Bell experiment setup appears
to be a good candidate to settle this issue, which is  
one of the basic unanswered question of quantum game theory.

To study quantum strategies in games of incomplete information,
we develop a formalism of game theory based on multi-sector probability matrix. 
We then analyze
a game of incomplete information which is an extension
of the well known game of Battle of Sexes and find the classical and the
quantum Bayesian Nash equilibria. We find two distinct effects of quantum
entanglement in games of incomplete information: \textit{pseudo-classical
distortion} and \textit{quantum nonlocality}. 
These two effects, in fact,
has been already identified as two separate correction terms in the payoff functions
in a previous study of games with complete information \cite{CT06}.  
It has been found there, however, that the pseudo-classical term, which can be simulated
classically, tends to overshadow the subtle effect of quantum nonlocality term.
It is shown, in this work, that the purely quantum element of quantum game strategy
can be unambiguously separated in a proper setup utilizing Bell inequality, 
and that setup is exactly found in Harsanyi's theory of games with incomplete information.  
\\

\section{Joint Probability Formalism of Incomplete Information Game} 
We start by formulating game strategies in terms of joint probabilities which do not, 
in general, factorize into individual player strategies \cite{IC07}.
Consider a system consisting of two players, Alice and Bob, who are
to  {play two-strategy games}, that is, 
to make selection from respective dichotomic choices, which
we label as $A = 0$ or $1$ for Alice, and $B = 0$ or $1$ for Bob.
The players are assumed to be autonomous decision makers interested 
in increasing their respective utility functions, or {\it payoffs} $\Pi_{Alice}$ 
and $\Pi_{Bob}$.
Game theory tries to answer the question what the stable pattern of selections
are after sufficient repetitions of game plays.
In the game theory, both payoffs $\Pi_{Alice}$ and $\Pi_{Bob}$ are 
functions of $A$ and $B$ at the same time.  In general, there is no unilateral 
optimal choice for neither players.

In determining the form of payoffs, we assume that not all information
necessary to specify payoff functions are known to players.
Following Harsanyi \cite{HA67}, 
we represent this unknown elements of the game by the concept of
{\it player type};
Both players comes into the play in one of two types
denoted by $a = 0$, $1$ for Alice, and $b = 0$, $1$ for Bob, and 
payoffs are uniquely determined only after determination of types.
Specifically, when Alice in type $a$ mode makes her move $A$ and
Bob in $b$ makes his move $B$, we assign real numbers $M^{[ab]}_{AB}$
for Alice's payoff, and $L^{[ab]}_{AB}$ for Bob's.  With varying indices $a$, $b$,
$A$ and $B$, both $M^{[ab]}_{AB}$ and $M^{[ab]}_{AB}$ form payoff matrices.

After sufficient run of repeated game play, 
the pattern of the play is specified by the joint probability
$P^{[ab]}_{AB}$ which represents the fraction of plays in which the move of Alice of
type $a$ is $A$, and that of Bob of type $b$, $B$. The average payoffs for Alice and Bob
of respective types $a$ and $b$ are  given by
\begin{eqnarray}
\label{ee01}
\Pi^{[ab]}_{Alice} = \sum_{A,B} M^{[ab]}_{AB} P^{[ab]}_{AB} ,
\quad
\Pi^{[ab]}_{Bob} = \sum_{A,B} L^{[ab]}_{AB} P^{[ab]}_{AB}  .
\end{eqnarray}
As probabilities, $P^{[ab]}_{AB}$ satisfy the relations
$\sum_{A,B} P^{[ab]}_{AB} = 1$
for any given types of players $a$ and $b$.
If we further assume that the types of the players  at each turn of play
is determined randomly (by Nature's move) with probabilities $S^{[a]}$ and $T^{[b]}$,
we obtain the total average payoffs  in the forms
\begin{eqnarray}
\label{ee02}
\Pi_{Alice} \!\!=\! \sum_{a,b} S^{[a]} T^{[b]} \Pi^{[ab]}_{Alice} ,
\ 
\Pi_{Bob} \!=\!  \sum_{a,b} S^{[a]} T^{[b]} \Pi^{[ab]}_{Bob} .
\end{eqnarray}
Central to the theory of game with incomplete information is the
assumption of {\it local knowledge of player types}, which postulates
that the type of a player at each turn of the play
is known only to herself (himself) and not to the other player.  
The statistical distributions of types $S^{[a]}$, $T^{[b]}$ are treated as
common knowledge.
It then follows that the pattern of play, or the {\it strategy} of Alice, 
which we assign symbol $\alpha$,
has to be determined only by the knowledge of $a$, but not with $b$.  Likewise
strategy of Bob $\beta$ can depend on his type $b$ but not on Alice's $a$.
Since the strategies of Alice and Bob jointly determine the joint probability of play,
we can express the assumption of locality of player type as
\begin{eqnarray}
\label{ee03}
P^{[ab]}_{AB} = P^{[ab]}_{AB} (\alpha^{[a]}, \beta^{[b]}) .
\end{eqnarray}
In traditional game theory, which has exclusively considered strategy based on classical
resources, the joint probability is given by the product of individual probabilities
as
\begin{eqnarray}
\label{ee04}
P^{[ab]}_{AB} = P^{[a]}_{A} Q^{[b]}_{B} ,
\end{eqnarray}
where $P^{[a]}_{A}$ represents the probability of Alice of type $a$ selecting the move $A$,
and $Q^{[b]}_{B}$ the probability of Bob of type $b$ selecting the move $B$.
In this case, we can identify the $P^{[a]}_{A}$ itself as 
Alice's strategy $\alpha^{[a]}$ and
$Q^{[b]}_{B}$ itself  as Bob's strategy  $\beta^{[b]}$, and no distinction between 
strategy and individual probability is necessary.
To generate desired strategies, players 
need access to devices that can generate probability distribution, such as dices.
If, on the other hand, we are to consider strategy based on quantum resources, 
we can construct joint probability out of individual strategies that correspond
to the individual actions on Hilbert space vector. 
For definiteness, we adopt the
quantum strategy based on Schmidt decomposition \cite{IT07}, that is known to 
cover \textit{entire $2\times 2$ dimensional Hilbert space},
which is given by
\begin{eqnarray}
\label{ee05}
P^{[ab]}_{AB}(\alpha^{[a]}, \beta^{[b]}) 
= \left| {\left<AB \right| U_{\alpha^{[a]}} V_{\beta^{[b]}} \left| \Phi_{\gamma \phi} \right>} 
   \right| ^2,
\end{eqnarray}
where the ``initial'' state $ \left| \Phi_{\gamma \phi} \right>$ residing 
on $2 \times 2$ dimensional Hilbert space is given by 
\begin{eqnarray}
\label{ee06}
\left| \Phi_{\gamma \phi} \right> 
=   \cos\frac{\gamma}{2} \left| 00 \right>  + e^{i\phi} \sin\frac{\gamma}{2} \left| 11 \right> ,
\end{eqnarray}
and individual rotations $U_\alpha$ and $V_\beta$ in $2$ dimensional subspaces, 
which are now identified as individual strategies, are given by
\begin{eqnarray}
\label{ee07}
&&
U_\alpha \left| 0 \right> = \cos\frac{\alpha}{2}  \left| 0 \right>
+  \sin\frac{\alpha}{2} \left| 1 \right>,
\nonumber \\
&&
U_\alpha \left| 1 \right> = -\sin\frac{\alpha}{2}  \left| 0 \right>
+  \cos\frac{\alpha}{2} \left| 1 \right>,
\nonumber \\
&&
V_\beta \left| 0 \right> = \cos\frac{\beta}{2}  \left| 0 \right>
+  \sin\frac{\beta}{2} \left| 1 \right>,
\\ \nonumber
&&
V_\beta \left| 1 \right> = -\sin\frac{\beta}{2}  \left| 0 \right>
+  \cos\frac{\beta}{2} \left| 1 \right>.
\end{eqnarray}
Note that, in addition to individual strategy variables $\alpha$ and $\beta$,
which are defined within the range $[0,\pi]$,
that are respectively controlled by Alice and Bob, there appear
two more variables $\gamma\in [0,\pi]$ and $\phi\ [0,\pi]$ 
``from nowhere'', as a result of
the requirement that strategies be described by Hilbert space vectors.
A natural interpretation of these new variables is that they belong to a 
third person, {\it the coordinator} of the game\cite{CT06}.
There are alternative choices of quantum strategies \cite{CT06,NT04} 
than the one given by (\ref{ee05})-(\ref{ee07}), but they do not
change our main conclusion, as long as entire Hilbert space is exhausted,
and thus all possible quantum joint probabilities are included.
The quantum joint probability (\ref{ee05}) can be realized, for example, by
the coordinator 
first generating two $z$-axis-polarized spins in entangled state (\ref{ee06}), 
then Alice and Bob obtaining one spin each, and performing spin 
rotations and their subsequent measurement
along $z$-axis, or equivalently, just measuring spins along properly rotated axes. 

If the initial state is prepared disentangled, for example in $\gamma=0$ state,
the quantum strategy (\ref{ee05}) is simply reduced to classical strategy (\ref{ee04})
with identification 
\begin{eqnarray}
\label{ee08}
P^{[a]}_A= |\left< A\right| U_{\alpha^{[a]}} \left| 0 \right>|^2 ,
\quad
Q^{[b]}_B= |\left< B\right| V_{\beta^{[b]}} \left| 0 \right>|^2 ,
\end{eqnarray}
which means that we have replaced usual dice by quantum spin systems
that act exactly as classical dices, albeit with far greater cost.

The payoffs are now the functions of strategy variables $\alpha$ and $\beta$,
and also of coordinator variables $\gamma$ and $\phi$;
\begin{eqnarray}
\label{ee09}
&&\Pi_{Alice}= \Pi_{Alice}(\alpha,\beta; \gamma,\phi),
\nonumber \\
&&\Pi_{Bob}= \Pi_{Bob}(\alpha,\beta; \gamma,\phi).
\end{eqnarray}
Here, we have adopted the obvious shorthand notations 
$\alpha = (\alpha^{[1]},\alpha^{[2]})$ and $\beta = (\beta^{[1]},\beta^{[2]})$.
Once payoff functions are calculated as functions of strategies,
the solution of the game is given by constructing {\it Bayesian Nash equilibria} 
$(\alpha^{\star},\beta^{\star})$ which are
obtained from local maximum specified by
\begin{eqnarray}
\label{ee10a}
&&
\left. \frac{\partial}{\partial \alpha^{[a]}} \Pi_{Alice}(\alpha,\beta; \gamma,\phi)
\right|_{(\alpha^\star,\beta^\star)}= 0,
\nonumber \\ &&
\left.\frac{\partial}{\partial {\beta^{[b]}}}\Pi_{Bob}(\alpha,\beta; \gamma,\phi)
\right|_{(\alpha^\star,\beta^\star)}= 0. 
\end{eqnarray}
If the payoffs do not have maxima as functions of $\alpha$ and $\beta$,
{\it classical pure Nash equilibria} emerge as the ``edge'' solutions;
\begin{eqnarray}
\label{ee11a}
\alpha^{[a]\star}=0,\  \beta^{[b]\star}=0
\  {\rm if}
\!\!\!\!\!\!\!\!\!\!\!\!&&
\frac{\partial}{\partial \alpha^{[a]}} \Pi_{Alice}(\alpha,\beta; \gamma,\phi)
 \!< 0,
\nonumber \\ &&
\frac{\partial}{\partial {\beta^{[b]}}}\Pi_{Bob}(\alpha,\beta; \gamma,\phi)
\!< 0,
\end{eqnarray}
\begin{eqnarray}
\label{ee12a}
\alpha^{[a]\star}=\pi,\  \beta^{[b]\star}=0
\ {\rm if}
\!\!\!\!\!\!\!\!\!\!\!\!&&
\frac{\partial}{\partial \alpha^{[a]}} \Pi_{Alice}(\alpha,\beta; \gamma,\phi)
\! > 0,
\nonumber \\ &&
\frac{\partial}{\partial {\beta^{[b]}}}\Pi_{Bob}(\alpha,\beta; \gamma,\phi)
\! < 0, 
\end{eqnarray}
\begin{eqnarray}
\label{ee13a}
\alpha^{[a]\star}=0,\  \beta^{[b]\star}=\pi
\ {\rm if}
\!\!\!\!\!\!\!\!\!\!\!\!&&
\frac{\partial}{\partial \alpha^{[a]}} \Pi_{Alice}(\alpha,\beta; \gamma,\phi)
\! < 0,
\nonumber \\ &&
\frac{\partial}{\partial {\beta^{[b]}}}\Pi_{Bob}(\alpha,\beta; \gamma,\phi)
\! > 0,
\end{eqnarray}
\begin{eqnarray}
\label{ee14a}
\alpha^{[a]\star}=\pi,\  \beta^{[b]\star}=\pi
\ {\rm if}
\!\!\!\!\!\!\!\!\!\!\!\!&&
\frac{\partial}{\partial \alpha^{[a]}} \Pi_{Alice}(\alpha,\beta; \gamma,\phi)
\! > 0,
\nonumber \\ &&
\frac{\partial}{\partial {\beta^{[b]}}}\Pi_{Bob}(\alpha,\beta; \gamma,\phi)
\! > 0.
\end{eqnarray}
In all cases, the Bayesian Nash payoffs are obtained as
\begin{eqnarray}
\label{ee15a}
&&
\Pi_{Alice}^\star(\gamma,\phi)= \Pi_{Alice}(\alpha^\star,\beta^\star; \gamma,\phi),
\nonumber \\ &&
\Pi_{Bob}^\star(\gamma,\phi)= \Pi_{Bob}(\alpha^\star,\beta^\star; \gamma,\phi) ,
\end{eqnarray}
for all combinations of ${[ab]} = {[00]}$, $[10]$, $[01]$ and $[11]$.
Note that Bayesian Nash equilibria are defined for each fixed values for coordinator
variables. 
\\

\section{Extended Battle of Sexes Game} 

In this section, we analyze a particular example of
game with incomplete information, that shows the power of
quantum strategies in dramatical fashion. 
We now consider the following payoff matrices
\begin{eqnarray}
\label{ee16b}
M = \left(
\begin{array} {cc}
            \begin{array}{cc} +3 & 0 \cr 0 & +1 \end{array} 
          &\begin{array}{cc} -3 & 0 \cr 0 & -1 \end{array} \cr
            \begin{array}{cc} -3 & 0 \cr 0 & -1 \end{array}  
          &\begin{array}{cc} -1 & 0 \cr 0 & -3 \end{array} 
\end{array} 
         \right) ,
\nonumber \\
L =  \left(
\begin{array} {cc}
            \begin{array}{cc} +1 & 0 \cr 0 & +3 \end{array} 
          &\begin{array}{cc}-1 & 0 \cr 0 & -3 \end{array} \cr
            \begin{array}{cc} -1 & 0 \cr 0 & -3 \end{array}  
          &\begin{array}{cc}  -3 & 0 \cr 0 & -1 \end{array} 
\end{array}  
         \right) .
\end{eqnarray}
Here,  $2 \times 2$ blocks represent payoff matrices for fixed player types,
$[ab] = [00]$, $[01]$, $[10]$ and $[11]$
from top-left to bottom right, namely
\begin{eqnarray}
\label{ee17b}
M = 
\left(\!\! \begin{array}{cc}
             M^{[00]} & M^{[01]} \cr
             M^{[10]} & M^{[11]}
\end{array} \!\!\right) ,
\ 
L = 
\left(\!\! \begin{array}{cc}
             L^{[00]} & L^{[01]} \cr
             L^{[10]} & L^{[11]}
\end{array} \!\!\right) .
\end{eqnarray}
We assume the ``democratic'' mixture of two types,
$S^{[0]}=S^{[1]}=1/2$ and $T^{[0]}=T^{[1]}=1/2$.
The payoffs are given, in terms of joint probabilities $P^{[ab]}_{AB}$ as
\begin{eqnarray}
\label{ee18b}
&&\!\!\!\!
\Pi_{Alice} = 
\frac{3}{4} ( { P^{[00]}_{00} - P^{[10]}_{00} - P^{[01]}_{00} - P^{[11]}_{11} } )
\nonumber \\
&&\qquad\!\!
+\frac{1}{4}  ( { P^{[00]}_{11} - P^{[10]}_{11} - P^{[01]}_{11} - P^{[11]}_{00} } ),
\nonumber \\ 
&&\!\!\!\!
\Pi_{Bob} = 
 \frac{1}{4} ( { P^{[00]}_{00} - P^{[10]}_{00} - P^{[01]}_{00} - P^{[11]}_{11} } )
\\ \nonumber
&&\qquad\!\!
+\frac{3}{4}  ( { P^{[00]}_{11} - P^{[10]}_{11} - P^{[01]}_{11} - P^{[11]}_{00} } ).
\end{eqnarray}
It is easily seen that the two terms of $\Pi_{Alice}$ and those of $\Pi_{Bob}$
are identical apart from the different weights.
They are both made up of $P^{[ab]}_{AB}$ of four 
different type combinations $[ab]$.  
This is so by design, which soon becomes evident in the followings. 
The factors $3/4$ and $1/4$ are, of course, the result of our specific choice of
numbers $\pm 3$ and $\pm 1$ in the entries of payoff matrices $M$ and $L$, and 
any other positive numbers will leave our analysis essentially unchanged.
The game in ``main sector'' that is played by type $a=0$ Alice and type $b=0$ Bob
is nothing but usual Battle of Sexes game.  If both players are limited within this sector,
there are two obvious pure Nash equilibria, $(A^\star,B^\star) = (0,0)$ 
and $(A^\star,B^\star) = (1,1)$, or
equivalently, $(\alpha^{[0]^\star},\beta^{[0]^\star})=(0,0)$ and 
$(\alpha^{[0]^\star},\beta^{[0]^\star})=(\pi,\pi)$.  The former solution is
advantageous to Alice and the latter to Bob as being evident from
the payoffs $(\Pi^{[00]\star}_{Alice}, \Pi^{[00]\star}_{Bob}) = (3,1)$ for the former 
and $(\Pi^{[00]\star}_{Alice}, \Pi^{[00]\star}_{Bob}) = (1,3)$ for the latter. 
In the ``shadow sectors'' $[ab] = [10]$,
$[01]$ and $[11]$, the game table is that of Chicken Game. 
If both players are limited within each sectors, Nash equilibria are achieved
by $(A^\star,B^\star) = (1,0)$ and $(A^\star,B^\star) = (0,1)$, both of which
results in zero payoffs $(\Pi^{[ab]\star}_{Alice}, \Pi^{[ab]\star}_{Bob}) = (0,0)$.

For the full game with incomplete information, the lack of knowledge leaves the players
guessing on the type of other party, and they have to be content with settling with less payoffs
on average, in comparison with the case of full information above.  
For the calculation of full game, we define
\begin{eqnarray}
\label{ee19b}
&&\!\!\!\!
\Delta_{00} \equiv
 P^{[00]}_{00} - P^{[10]}_{00} - P^{[01]}_{00} - P^{[11]}_{11}  ,
\nonumber \\ 
&&\!\!\!\!
\Delta_{11} \equiv
 P^{[00]}_{11} - P^{[10]}_{11} - P^{[01]}_{11} - P^{[11]}_{00}  .
\end{eqnarray}
Obviously, we have Bayesian Nash equilibria when we have simultaneous maxima for
$\Delta_{00}$ and $\Delta_{11} $ as functions of $\alpha$ and $\beta$.
Explicit form for $\Delta_{00}$ is
\begin{eqnarray}
\label{ee20}
&&\!\!\!\!\!\!\!\!\!\!\!\!\!\!\!\!\!\!
\Delta_{00} = 
\cos^2\frac{\gamma}{2}
  \left( 
     \cos^2\frac{\alpha^{[0]}}{2} \cos^2\frac{\beta^{[0]}}{2}
    -\cos^2\frac{\alpha^{[1]}}{2} \cos^2\frac{\beta^{[0]}}{2}
  \right.
\nonumber \\&&\quad\!\!
  \left.  
    -\cos^2\frac{\alpha^{[0]}}{2} \cos^2\frac{\beta^{[1]}}{2}
    -\sin^2\frac{\alpha^{[1]}}{2} \sin^2\frac{\beta^{[1]}}{2}
  \right)
\nonumber \\ 
&&\!\!\!\!\!\!
+\sin^2\frac{\gamma}{2}
  \left( 
     \sin^2\frac{\alpha^{[0]}}{2} \sin^2\frac{\beta^{[0]}}{2}
    -\sin^2\frac{\alpha^{[1]}}{2} \sin^2\frac{\beta^{[0]}}{2}
  \right.
\nonumber \\&&\quad\!\!
  \left.  
    -\sin^2\frac{\alpha^{[0]}}{2} \sin^2\frac{\beta^{[1]}}{2}
    -\cos^2\frac{\alpha^{[1]}}{2} \cos^2\frac{\beta^{[1]}}{2}
  \right)
\nonumber \\
&&\!\!\!\!\!\!
+\frac{1}{4}\cos\phi\sin\gamma
  \left( 
     \sin\alpha^{[0]} \sin\beta^{[0]}
    -\sin\alpha^{[1]} \sin\beta^{[0]}
  \right.
\\ \nonumber
&&\qquad\qquad
  \left.  
    -\sin\alpha^{[0]} \sin\beta^{[1]}
    -\sin\alpha^{[1]} \sin\beta^{[1]}
  \right) .
\end{eqnarray}
The other quantity $\Delta_{11}$ is obtained by the simultaneous replacements 
$\alpha \to \pi-\alpha$ and $\beta \to \pi-\beta$,
or equivalently, by the replacement $\gamma \to \pi-\gamma$.
From these, we obtain the condition for Bayesian Nash equilibrium as
\begin{eqnarray}
\label{ee21}
&&\!\!\!\!\!\!
\sin\alpha^{[0]}(\cos\beta^{[0]}-\cos\beta^{[1]})
\nonumber \\ &&
-\cos\alpha^{[0]}(\sin\beta^{[0]}-\sin\beta^{[1]})\cos\phi\sin\gamma = 0 ,
\nonumber \\ 
&&\!\!\!\!\!\!
\sin\alpha^{[1]}(\cos\beta^{[0]}+\cos\beta^{[1]})
\nonumber \\ &&
-\cos\alpha^{[1]}(\sin\beta^{[0]}+\sin\beta^{[1]})\cos\phi\sin\gamma = 0 ,
\nonumber \\ 
&&\!\!\!\!\!\!
\sin\beta^{[0]}(\cos\alpha^{[0]}-\cos\alpha^{[1]})
\nonumber \\ &&
-\cos\beta^{[0]}(\sin\alpha^{[0]}-\sin\alpha^{[1]})\cos\phi\sin\gamma = 0 ,
\\ \nonumber
&&\!\!\!\!\!\!
\sin\beta^{[1]}(\cos\alpha^{[0]}+\cos\alpha^{[1]})
\\ \nonumber&&
-\cos\beta^{[1]}(\sin\alpha^{[0]}+\sin\alpha^{[1]})\cos\phi\sin\gamma = 0 .
\end{eqnarray}
The classical game is obtained as the limit of no entanglement, $\gamma=0$,
for which we recover the separability of probabilities, 
$P^{[ab]}_{AB}=P^{[a]}_{A} Q^{[b]}_{B}$.
There are eight sets of Bayesian Nash equilibria found in this game, 
all supporting the {\it break-even payoffs} 
\begin{eqnarray}
\label{ee22}
\Pi^\star_{Alice}=0,\quad \Pi^\star_{Bob} = 0 .
\end{eqnarray}
They are 
\begin{eqnarray}
\label{ee23}
&&\!\!\!\!\!\!\!\!\!\!\!\!\!\!\!\!
1:
\alpha^{[0]}=0, \ \alpha^{[1]}=0, \ \beta^{[0]}={\rm arbitrary}, \ \beta^{[1]}=\pi  ,
\nonumber \\
&&\!\!\!\!\!\!\!\!\!\!\!\!\!\!\!\!
2:
\alpha^{[0]}=\pi, \ \alpha^{[1]}=\pi, \ \beta^{[0]}={\rm arbitrary}, \ \beta^{[1]}=0  ,
\nonumber \\
&&\!\!\!\!\!\!\!\!\!\!\!\!\!\!\!\! 
3:
\alpha^{[0]}=0, \ \alpha^{[1]}=\pi, \ \beta^{[0]}=0, \ \beta^{[1]}={\rm arbitrary}  ,
\nonumber \\
&&\!\!\!\!\!\!\!\!\!\!\!\!\!\!\!\!
4: 
\alpha^{[0]}=\pi, \ \alpha^{[1]}=0, \ \beta^{[0]}=\pi, \ \beta^{[1]}={\rm arbitrary},
\\ \nonumber
&&\!\!\!\!\!\!\!\!\!\!\!\!\!\!\!\!
5: 
\alpha^{[0]}={\rm arbitrary}, \ \alpha^{[1]}=\pi, \ \beta^{[0]}=0, \ \beta^{[1]}=0  ,
\\ \nonumber
&&\!\!\!\!\!\!\!\!\!\!\!\!\!\!\!\!
6: 
\alpha^{[0]}={\rm arbitrary}, \ \alpha^{[1]}=0, \ \beta^{[0]}=\pi, \ \beta^{[1]}=\pi  ,
\\ \nonumber
&&\!\!\!\!\!\!\!\!\!\!\!\!\!\!\!\!
7: 
\alpha^{[0]}=0, \ \alpha^{[1]}={\rm arbitrary}, \ \beta^{[0]}=0, \ \beta^{[1]}=\pi  ,
\\ \nonumber
&&\!\!\!\!\!\!\!\!\!\!\!\!\!\!\!\! 
8:
\alpha^{[0]}=\pi, \ \alpha^{[1]}={\rm arbitrary}, \ \beta^{[0]}=\pi, \ \beta^{[1]}=0  .
\end{eqnarray}

There is a deeper reason for the fact that the Bayesian Nash equilibria for this game
only gives zero payoffs and nothing more: That is exactly the Bell inequalities.
In our setting of $2 \times 2$ game, Bell inequalities are the relations among 
joint probabilities for Alice and Bob each being capable of turning up
in two types $a=0$, $1$ and $b=0$, $1$.  If each player choose her/his probability
only with the knowledge of her/his own type, but not of other player, 
a set of inequalities can be proven.  Since this condition is exactly 
the type-locality assumption we have postulated in the game of incomplete information,
it is reasonable that we expect Bell inequalities to be satisfied in our settings.
A specifically relevant ones in our case are the inequalities
first proven by Cereceda \cite{CE01} which read
\begin{eqnarray}
\label{ee24}
&&\!\!\!\!
P^{[00]}_{00} - P^{[10]}_{00} - P^{[01]}_{00} - P^{[11]}_{11}  \le 0 ,
\nonumber \\ 
&&\!\!\!\!
P^{[00]}_{11} - P^{[10]}_{11} - P^{[01]}_{11} - P^{[11]}_{00}  \le 0 .
\end{eqnarray}
There are 64 Cereceda inequalities obtainable by renaming of superscripts $[ab]$
and subscripts $AB$, that can be divided into 16 quartets.  Each quartet
sums up to give a single CHSH inequality, and can be regarded as a set of 
``elementary'' pieces of a CHSH inequality. 
Each Cereceda inequality contains all four combinations of types $[ab]$.
It is shown by Fine \cite{FI82} that the Bell inequality breaking occurs if and only if 
the assumption of \textit{factorizability of joint probabilities}, (\ref{ee04}), is violated.  
Since LHSs of the Cereceda inequality, (\ref{ee24}) are nothing other than 
$\Delta_{00}$ and $\Delta_{11}$, we always have
\begin{eqnarray}
\label{ee25}
\Pi_{Alice} \le 0, \quad
\Pi_{Bob} \le 0 .
\end{eqnarray}
We can now see that, with the Bayesian Nash payoffs (\ref{ee22}), both players are
getting maximum payoffs mathematically possible under the assumption of 
type-locality.

If players are allowed to share quantum objects, it is possible to have nonlocal strategies,
and it is expected that the classical limit of payoffs imposed by Cereceda inequality
can be exceeded.

There are, however, limits on the amount of quantum breaking of Bell inequalities.
According to Cirel'son \cite{TS80}, the nonlocality supplied by quantum mechanics can
break the Cereceda-Bell inequality up to the following amount;
\begin{eqnarray}
\label{ee26}
&&\!\!\!\!
P^{[00]}_{00} - P^{[10]}_{00} - P^{[01]}_{00} - P^{[11]}_{11}  \le \frac{\sqrt{2}-1}{2} ,
\nonumber \\ 
&&\!\!\!\!
P^{[00]}_{11} - P^{[10]}_{11} - P^{[01]}_{11} - P^{[11]}_{00}  \le \frac{\sqrt{2}-1}{2} .
\end{eqnarray}
which should limits the possible payoffs to
\begin{eqnarray}
\label{ee27}
\Pi_{Alice} \le \frac{\sqrt{2}-1}{2}, 
\quad
\Pi_{Bob} \le \frac{\sqrt{2}-1}{2} ,
\end{eqnarray}
even with the quantum strategies.

As is well known, Bell inequalities are, in general, maximally broken when the 
quantum entanglement is largest.  For the quantum strategy given by
(\ref{ee05}) and (\ref{ee06}), that corresponds to the case of $\gamma=\pi/2$
and $\phi=0$.  
In this case, Bayesian Nash condition (\ref{ee21}) becomes 
\begin{eqnarray}
\label{ee28}
&&\!\!\!\!
\sin(\alpha^{[0]}-\beta^{[1]}) - \sin(\alpha^{[0]}-\beta^{[0]}) = 0 ,
\nonumber \\ 
&&\!\!\!\!
\sin(\alpha^{[1]}-\beta^{[1]}) + \sin(\alpha^{[1]}-\beta^{[0]}) = 0 ,
\nonumber \\ 
&&\!\!\!\!
\sin(\alpha^{[1]}-\beta^{[0]}) - \sin(\alpha^{[0]}-\beta^{[0]}) = 0 ,
\\ \nonumber
&&\!\!\!\!
\sin(\alpha^{[1]}-\beta^{[1]}) + \sin(\alpha^{[0]}-\beta^{[1]}) = 0 .
\end{eqnarray}
From this condition, we can identify a single set of quantum Bayesian Nash equilibria
for the case of $\cos\phi\sin\gamma=1$ as
\begin{eqnarray}
\label{ee29}
&&
\beta^{[0]\star}-\alpha^{[0]\star}=\frac{\pi}{4} ,
\nonumber \\&&
\beta^{[1]\star}-\alpha^{[0]\star}=\frac{3\pi}{4} ,
\\ \nonumber&&
\alpha^{[1]\star}-\beta^{[0]\star}=\frac{5\pi}{4} .
\end{eqnarray}
Since there are only three constraints on four quantities,
Bayesian Nash we have is a continuous set.
For this set of values, we have
\begin{eqnarray}
\label{ee30}
\Delta_{00} = \Delta_{11} = \cos^2\frac{\pi}{8}-3\sin^2\frac{\pi}{8},
\end{eqnarray}
which immediately leads to the quantum Bayesian Nash payoffs
\begin{eqnarray}
\label{ee31}
\Pi^\star_{Alice} =\frac{\sqrt{2}-1}{2}, 
\quad
\Pi^\star_{Bob} =\frac{\sqrt{2}-1}{2} .
\end{eqnarray}
In this quantum case again, both players are getting maximal payoffs allowable
under the Cirel'son limit (\ref{ee26}).
Note again that positive payoffs are never possible under classical strategies
even with correlations, for example, cheap-talk and altruism, and it is a signature of
nonclassical correlation inherent in quantum strategies.
%
\\

\section{Pseudo-classical and Quantum Interference Components}
In order to examine the physical contents of the quantum strategies in detail,
we define the single player probabilities
\begin{eqnarray}
\label{ee32}
p_{A}^{[a]}
= \left| {\left<A | U_{\alpha^{[a]}} | 0 \right>} \right| ^2 ,
\quad
q_{B}^{[b]}
= \left| {\left<B \right| U_{\beta^{[b]}} \left| 0 \right>} \right| ^2 ,
\end{eqnarray}
with which, we express the joint probability
\begin{eqnarray}
\label{ee33}
&&\!\!\!\!\!\!\!\!\!\!\!\!
P_{AB}^{[ab]} 
=  \cos^2\frac{\gamma}{2} p_A^{[a]} q_B^{[b]}
  +\sin^2\frac{\gamma}{2} p_{\bar A}^{[a]} q_{\bar B}^{[b]}
\nonumber \\ &&\qquad
  +(-)^{A+B}
  \cos\phi \sin\gamma\sqrt{p_A^{[a]} p_{\bar A}^{[a]} q_B^{[b]} q_{\bar B}^{[b]}} .
\end{eqnarray}
Here, the notation ${\bar 0} = 1$, ${\bar 1} = 0$ is used.
The sector payoffs $\Pi^{ab}$ take the form
\begin{eqnarray}
\label{ee34}
&&\!\!\!\!\!\!\!\!\!\!\!\!\!\!\!\!\!\!\!\!\!\!
\Pi^{[ab]}_{Alice} = \sum_{A,B} \,(\, \cos^2\frac{\gamma}{2} M^{[ab]}_{AB} 
+ \sin^2\frac{\gamma}{2} L^{[ab]}_{AB} \,)\,  p^{[a]}_{A} q^{[b]}_{B} 
\nonumber \\ &&\!\!\!\!
+ \cos\phi\sin\gamma   
\sqrt{p_0^{[a]} p_1^{[a]} q_0^{[b]} q_1^{[b]}} \sum_{A,B} (-)^{A+B} M^{[ab]}_{AB},
\nonumber \\ 
&&\!\!\!\!\!\!\!\!\!\!\!\!\!\!\!\!\!\!\!\!\!\!
\Pi^{[ab]}_{Bob} = \sum_{A,B} \,(\, \cos^2\frac{\gamma}{2} L^{[ab]}_{AB} 
+ \sin^2\frac{\gamma}{2} M^{[ab]}_{AB} \,)\,  p^{[a]}_{A} q^{[b]}_{B} 
\nonumber \\ &&\!\!\!\!
+ \cos\phi\sin\gamma   
\sqrt{p_0^{[a]} p_1^{[a]} q_0^{[b]} q_1^{[b]}} \sum_{A,B} (-)^{A+B} L^{[ab]}_{AB} ,
\end{eqnarray}
where we have used the symmetry property $L^{[ab]}_{AB}=M^{[ab]}_{{\bar A}{\bar B}}$
of our game.
In this form, we clearly see that both payoffs are composed of two components.
First component, formerly termed as classical family \cite{CT06} 
represents essentially classical payoff coming 
from ``altruistic'' modification of the game matrix \cite{CH03,CH05}. 
Even with this modification, the payoffs are still constructed 
from factorizable probabilities $p^{[a]}_A q^{[b]}_B$, 
and therefore, payoffs will never exceed the limit (\ref{ee25}).
This leaves the second  component, previously known as interference term, 
as the sole source of truly quantum gain in the payoff  
that is achieved through the Bell inequality breakings.  

In hindsight, it should have been naturally expected that the probabilities
generated from ``successful'' quantum strategies shall break some form of Bell
inequalities, since extra quantum gains obtained from such strategies should be, 
by definition, the result of breakdown of the assumption of factorizability, (\ref{ee04}).
However, in order to establish a {\it direct link} between the Bell inequality breaking and
the extra quantum gain in game's payoff, it is necessary to  have
a proper game theoretic setup,
and that is exactly what we have shown here with the game of incomplete information.
It is rather miraculous that essentially identical setup has been
conceived contemporaneously in two separate disciplines as Harsanyi's game theory 
and Bell's quantum measurement theory.  
It seems possible that future investigation may reveal a hidden intellectual thread
between the two.   It could also be that they originate from a common 
mid-twentieth century \textit{Zeitgeist}.
\\

\section{Extensions to Many Player Games}
The following inequality 
is shown to hold by Cereceda\cite{CE04} for $2\times 2 \times 2$ system, that is
a system with three spins measured by three observers,
Alice, Bob and Chris, each equipped with detector capable of performing
spin projection measurement along two possible directions. 
\begin{eqnarray}
\label{ee35}
&&\!\!\!\!\!\!\!\!
P^{[000]}_{000} -P^{[100]}_{000} - P^{[010]}_{000} 
- P^{[001]}_{000} - P^{[111]}_{111}   \le 0 ,
\nonumber \\
&&\!\!\!\!\!\!\!\!
P^{[000]}_{111} -P^{[100]}_{111} - P^{[010]}_{111} 
- P^{[001]}_{111} - P^{[111]}_{000}   \le 0 .
\end{eqnarray}
It is easy to conceive a $2\times 2 \times 2$ game that shows quantum gain
using this inequality.  Following type of game matrix will do;
\begin{eqnarray}
\label{ee36}
&&
\{M^{[0bc]}_{0BC}\} = \left(
\begin{array} {cc}
            \begin{array}{cc}  +3 & 0 \cr 0 & 0 \end{array} 
          &\begin{array}{cc}  -3 & 0 \cr 0 & 0 \end{array} \cr
            \begin{array}{cc}  -3 & 0 \cr 0 & 0 \end{array}  
          &\begin{array}{cc} 0 & 0 \cr 0 & 0 \end{array} 
\end{array}  \right),
\nonumber \\&&
\{L^{[0bc]}_{0BC}\} = \left(
\begin{array} {cc}
            \begin{array}{cc} +1 & 0 \cr 0 & 0 \end{array} 
          &\begin{array}{cc} -1 & 0 \cr 0 & 0 \end{array} \cr
            \begin{array}{cc} -1 & 0 \cr 0 & 0 \end{array}  
          &\begin{array}{cc} 0 & 0 \cr 0 & 0 \end{array} 
\end{array} \right),
\nonumber \\
&&
\{M^{[0bc]}_{1BC}\} = \left(
\begin{array} {cc}
            \begin{array}{cc} 0 & 0 \cr 0 & +1 \end{array} 
          &\begin{array}{cc} 0 & 0 \cr 0 & -1 \end{array} \cr
            \begin{array}{cc} 0 & 0 \cr 0 & -1 \end{array}  
          &\begin{array}{cc} 0 & 0 \cr 0 & 0 \end{array}  
\end{array} \right),
\nonumber \\&&
\{L^{[0bc]}_{1BC}\} = \left(
\begin{array} {cc}
            \begin{array}{cc}  0 & 0 \cr 0 & +3 \end{array} 
          &\begin{array}{cc} 0 & 0 \cr 0 & -3 \end{array} \cr
            \begin{array}{cc} 0 & 0 \cr 0 & -3 \end{array}  
          &\begin{array}{cc} 0 & 0 \cr 0 & 0 \end{array} 
\end{array}  \right),
\nonumber \\
&&
\{M^{[1bc]}_{0BC}\} = \left(
\begin{array} {cc}
            \begin{array}{cc} -3 & 0 \cr 0 & 0 \end{array} 
          &\begin{array}{cc} 0 & 0 \cr 0 & 0 \end{array} \cr
            \begin{array}{cc} 0 & 0 \cr 0 & 0 \end{array}  
          &\begin{array}{cc} -1 & 0 \cr 0 & 0 \end{array}
\end{array} \right),
\nonumber \\&&
\{L^{[1bc]}_{0BC}\} = \left(
\begin{array} {cc}
            \begin{array}{cc} -1 & 0 \cr 0 & 0 \end{array} 
          &\begin{array}{cc} 0 & 0 \cr 0 & 0 \end{array} \cr
            \begin{array}{cc} 0 & 0 \cr 0 & 0 \end{array}  
          &\begin{array}{cc} -3 & 0 \cr 0 & 0 \end{array} 
\end{array} \right),
\\ \nonumber
&&
\{M^{[1bc]}_{1BC}\} = \left(
\begin{array} {cc}
            \begin{array}{cc} 0 & 0 \cr 0 & -1 \end{array} 
          &\begin{array}{cc} 0 & 0 \cr 0 & 0 \end{array} \cr
            \begin{array}{cc} 0 & 0 \cr 0 & 0 \end{array}  
          &\begin{array}{cc} 0 & 0 \cr 0 & -3 \end{array} 
\end{array} \right) ,
\\ \nonumber&&
\{L^{[1bc]}_{1BC}\} = \left(
\begin{array} {cc}
            \begin{array}{cc}  0 & 0 \cr 0 & -3 \end{array} 
          &\begin{array}{cc} 0 & 0 \cr 0 & 0 \end{array} \cr
            \begin{array}{cc} 0 & 0 \cr 0 & 0 \end{array}  
          &\begin{array}{cc} 0 & 0 \cr 0 & -1 \end{array} 
\end{array} \right) .           
\end{eqnarray}
Further geralization of this result to $2\times 2 \times ... \times 2$ system is
\begin{eqnarray}
\label{ee37}
&&\!\!\!\!\!\!\!\!\!\!
P^{[00...0]}_{00...0}
- P^{[100...0]}_{00...0} - P^{[010...0]}_{00...0} -...
\nonumber \\&&\qquad\qquad\qquad\ 
- P^{[00...01]}_{00...0} - P^{[11...1]}_{11..1}  \le 0 ,
\nonumber \\
&&\!\!\!\!\!\!\!\!\!\!
P^{[00...0]}_{11...1}
- P^{[100...0]}_{11...1} - P^{[010...0]}_{11...1} -...
\nonumber \\&&\qquad\qquad\qquad\ 
- P^{[00...01]}_{11...1} - P^{[11...1]}_{00...0}  \le 0 .
\end{eqnarray}
It should again be easy to formulate a multi-party game based on this general Cereceda
inequality. 

Although current set of examples of fully solvable quantum game of incomplete information is
of rather special sort, having very sparse nonzero elements, it shows the
two different novel aspects of quantum strategy that are not
present in classical counterpart in very clear fashion. 
The relation between purely quantum gain in the payoff and Bell inequality breaking
is indeed striking.
It is obvious that the same effects should persist in more general quantum games
albeit with less clearly discernible form. 

In summary, we have shown that there is a genuine advantage in quantum strategy
that is not accessible by classical resources, and that advantage
is to be found most clearly and unambiguously
in a refined settings of Harsanyi's games of incomplete information.  

\section*{Acknowledgment}
This work has been partially supported by 
the Grant-in-Aid for Scientific Research of  Ministry of Education, 
Culture, Sports, Science and Technology, Japan
under the Grant numbers 18540384 and 18.06330.
 

\end{document}